\documentclass[epj]{svjour}
\input epsf

\newcounter{append}
\newcommand\Appendix[1]{
 \setcounter{equation}{0}
 \setcounter{subsection}{0}
 \addtocounter{append}{1}
 \def\theappend{\Alph{append}}
 \def\thesection{\theappend}
 \renewcommand{\theequation}{\theappend.\arabic{equation}}
  \section*{Appendix: #1}}
\newcommand{\bc}{\begin{center}}
\newcommand{\ec}{\end{center}}
\newcommand{\be}{\begin{equation}}
\newcommand{\ee}{\end{equation}}
\newcommand{\ba}{\begin{array}}
\newcommand{\ea}{\end{array}}
\newcommand{\beqn}{\begin{eqnarray}}
\newcommand{\eeqn}{\end{eqnarray}}

\begin{document}

\title{Random quantum magnets with broad disorder distribution}

\author{D. Karevski\inst{1}, Y-C. Lin\inst{2}, H. Rieger\inst{3}, N. Kawashima\inst{4} and F. Igl\'oi\inst{5,6,1} }

\institute{
Laboratoire de Physique des Mat\'eriaux, Universit\'e Henri
Poincar\'e (Nancy 1), B.P. 239,\\ F-54506 Vand\oe uvre l\`es Nancy cedex, France \and
NIC, Forschungszentrum J\"ulich, 52425 J\"ulich, Germany \and
Theoretische Physik, Universit\"at des Saarlandes, 
     66041 Saarbr\"ucken, Germany \and
Department of Physics, Tokyo Metropolitan University, 192-0397 Tokyo, Japan \and   
Research Institute for Solid State Physics and Optics, 
H-1525 Budapest, P.O.Box 49, Hungary \and
Institute for Theoretical Physics,
Szeged University, H-6720 Szeged, Hungary
}

\date{January 16, 2001}
\abstract
{We study the critical behavior of Ising quantum magnets with broadly distributed random
couplings $(J)$, such that  $P(\ln J) \sim |\ln J|^{-1-\alpha}$, $\alpha>1$, for large
$|\ln J|$ (L\'evy flight statistics). For sufficiently broad distributions, $\alpha<\alpha_c$,
the critical behavior is controlled by a line of fixed points, where the critical exponents
vary with the L\'evy index, $\alpha$. In one dimension, with $\alpha_c=2$, we obtaind
several exact results through a mapping to surviving Riemann walks. In two dimensions
the varying critical exponents have been calculated by a numerical implementation of the
Ma-Dasgupta-Hu renormalization group method leading to $\alpha_c \approx 4.5$. Thus in the region
$2<\alpha<\alpha_c$, where the central limit theorem holds for $|\ln J|$
the broadness of the distribution is relevant for the 2d quantum Ising model.
}

\authorrunning{D. Karevski et al.}
\titlerunning{Random quantum magnets with broad disorder distribution}

\maketitle
\section{Introduction}
Quenched, i.e. time independent disorder may have a strong influence on quantum phase
transitions, which take place at zero temperature by varying a quantum control parameter, $\delta$,
which measures the strength of quantum fluctuations\cite{bhatt}.  The critical behavior of random
quantum magnets can be conveniently studied within the framework of a renormalization group (RG) scheme
introduced by Ma, Dasgupta and Hu\cite{mdh} and later applied by Fisher\cite{fisher,fisherxx} and
others\cite{2drg,lkir,ki}. In
this ``energy-space'' RG method the strongest bonds and transverse fields are successively decimated
out and other couplings/fields  are replaced by weaker ones generated by a perturbation calculation.
During the RG
procedure one keeps track of the energy scale, $\Omega$, which is the actual value of the
strongest coupling/transverse field; the average length scale, $\xi$, which is the correlation
length associated with the average equal-time correlation function; and the size
of the average local moment of a cluster, $\mu$, which is formed after decimating out
the couplings between the spins. For many systems - during renormalization - the disorder
grows without limits, thus as $\Omega \to 0$ typically a coupling/transverse field is
infinitely stronger or weaker than the neighboring one and the RG procedure becomes asymptotically
exact\cite{RGGR}. This type of critical behavior is controlled by an infinite
randomness fixed point (IRFP),  At an IRFP the physical
properties of the system are determined by the so called
{\it rare events}, which are samples, which occur with vanishing probability, but dominate the
average behavior. 

At an IRFP the scaling properties are unusual: they are fully characterized by three critical
exponents, denoted by $\psi$, $\phi$ and $\nu$ and defined as follows\cite{fisherA}. At
the IRFP the relevant time-scale, $t_r \sim 1/\Omega$, and length-scale, $\xi$, are related through:
\be
\ln t_r \sim \xi^{\psi}\;,
\label{psi}
\ee
thus scaling is strongly anisotropic and, as a consequence, the average critical dynamical correlations
are logarithmically slow\cite{riegerigloi97}. The dependence of the average magnetic moment per cluster
on the change of the energy scale as $\Omega_0 \to \Omega$ is asymptotically given by
\be
\mu \sim \left[\ln(\Omega_0/\Omega)\right]^{\phi}\;,
\label{phi}
\ee
from which the anomalous dimension of the bulk magnetization, $x_m$, in dimension, $d$, follows as:
\be
x_m=d-\phi \psi\;.
\label{xm}
\ee
We note that for bulk spins the average, equal-time correlations behave as $\sim r^{-2 x_m}$ at criticality,
whereas end-to-end correlations involve the corresponding surface exponent,
$x_m^s$, which is generally different form $x_m$. Finally, the third characteristic exponent of the IRFP,
$\nu$, is connected to
the asymptotic behavior of the average correlation length close to the critical point:
\be
\xi \sim |\delta|^{-\nu}\;.
\label{nu}
\ee
Prototype of a disordered quantum system with an IRFP is the random transverse-field Ising model
(RTIM) defined by the Hamiltonian:
\be
H=-\sum_{<i,j>} J_{ij} \sigma_i^x \sigma_j^x - \sum_i h_i \sigma_i^z\;.
\label{hamilton}
\ee
Here $\sigma_i^x,\sigma_i^z$ are Pauli matrices at site $i$, the nearest neighbor coupling constants,
$J_{ij}$, and the transverse fields, $h_i$, are independent random variables.

In one dimension (1d), where the topology of the lattice is invariant under renormalization,
Fisher\cite{fisher} has solved analytically the fixed-point RG equations
from which the position of the random critical point,
\be
\delta=\left[\ln h\right]_{\rm av}-\left[\ln J\right]_{\rm av}=0\;,
\label{delta}
\ee
and the values of the critical exponents follow. They are listed in Table 1.

In Eq.(\ref{delta}) and in the following $[ \dots ]_{\rm av}$ is used to denote averaging over
quenched disorder. The critical exponents in Table 1, except $\phi$, can also be derived by an
exact mapping to a random walk (RW) problem\cite{bigpaper}. A series of numerical
studies\cite{numerical} have confirmed the validity of the above RG and RW results.

\begin{table}
\caption{Critical exponents of the RTIM in one and two dimensions. In 1d exact results
for normal disorder are in the first row, whereas in the second row the corresponding
exponents are given for the L\'evy distribution
$(1<\alpha<2)$. In 2d the differences in the numerical
estimates are due to different disorder distributions and the truncation procedure used in the
numerical RG method.}
\label{TAB01}
\begin{tabular}{lrrrrr}
\hline\noalign{\smallskip}
 & $\Psi$ & $\Phi$ & $\nu$ & $x_m$  & $x_m^s$  \\
\hline\noalign{\smallskip}
$1d$ & $1/2$ & $\frac{1+\sqrt{5}}{2}$ & 2 & $\frac{3-\sqrt{5}}{4}$ & 1/2  \\
L\'evy & $1/\alpha$ &  & $\alpha/(\alpha-1)$ &  & 1/2  \\
\hline\noalign{\smallskip}
 & $0.42^{[5]}$ & $2.5^{[5]}$ & $1.072^{[5]}$ & $1.0^{[5]}$ &  \\
$2d$ & $0.5^{[6]}$ & $2.0^{[6]}$ & & $0.94^{[6]}$ &  \\
 & $0.6^{[7]}$ & $1.7^{[7]}$ & $1.25^{[7]}$ & $0.97^{[7]}$ &  \\
\noalign{\smallskip}\hline
\end{tabular}
\end{table}

Also in two dimensions the critical behavior of the RTIM is found to be controlled by an IRFP\cite{2drg}
and the corresponding critical exponents, as presented in Table 1, have been determined by
implementing numerically the RG procedure\cite{2drg,lkir,ki}. These RG results are consistent
with values obtained by quantum Monte Carlo simulations\cite{mc}.

One important question concerning the critical behavior of
random quantum systems is the domain of attraction of the IRFP for
different type of disorder distributions. Using an analogy with the RW
problem it is generally expected that the critical exponents of random
quantum systems are universal, provided the disorder is {\it i)} spatially
homogeneous, {\it ii)} not correlated, and {\it iii)} not broadly distributed.
Problems related to the first two conditions, i.e. the effect of inhomogeneous
or correlated disorder have already been studied in Ref.\cite{kjti}
and Ref.\cite{riegerigloi99}, respectively. In both cases the critical
properties for strong enough perturbations are
modified: they are governed by a line of IRFP-s, such that the
critical exponents are continuously varying functions of the
inhomogeneity/correlation parameters.

In the present paper we are going to release the third type of restriction
and study the effect of broad disorder distributions on the critical properties of
random quantum magnets. We consider the RTIM with the Hamiltonian in
Eq.(\ref{hamilton}) and keep in mind that at the IRFP it is the logarithm of the couplings
and the transverse fields  which follows a smooth probability distribution\cite{fisher}. Therefore
we use a parametrization
\beqn
J_{ij}&=& \Lambda^{\Theta_{ij}}\nonumber\\
h_i&=&h_0 \;.
\label{parametr}
\eeqn
where the exponents, $\Theta_{ij}$ are independent
random variables. They are taken from a broad distribution, $\pi(\Theta)$, such
that for large arguments they decrease
as, $\pi(\Theta) \sim |\Theta|^{-1-\alpha}$. We consider the region $ \alpha >1$, where
the $\kappa$-th moment of the distribution exists for $\kappa<\alpha$.
This type of distribution, which comes for in different domains of physics
and science\cite{bak} is usually called {\it L\'evy flight} or {\it Riemann walk} in the
discrete version its parameter, $\alpha$, is the L\'evy index.
Throughout this paper we use the following distribution:
$\pi(\Theta)=p {\alpha}  (1+\Theta)^{-1-\alpha}$ for $\Theta>0$ and.
$\pi(\Theta)=q {\alpha}  (1+|\Theta|)^{-1-\alpha}$ for $\Theta<0$, $p+q=1$. In one dimension
the quantum control parameter in Eq.(\ref{delta}) is given by  
$\delta=\ln h_{0}-(p-q)\ln \Lambda /(\alpha-1)$. We also use the
discretized
version (Riemann walk) of the above distribution, where $\Theta_i=\pm 1, \pm 2, \dots$ {\it etc.} 
and the normalization
$\alpha$ is replaced by $(\zeta(1+\alpha)-1)$, where $\zeta(x)$ denotes the Weierstrass
zeta function.

The structure of the paper is the following. In Section 2 the 1d version of the model is
considered and studied analytically (through a mapping to surviving Riemann walks) and
numerically. In Section 3 the 2d model is studied by a numerical implementation
of the RG procedure. Our results are discussed in the final Section, whereas some
details about persistence properties of L\'evy flights are presented in the Appendix.

\section{The one-dimensional problem}
\subsection{Exact results through a mapping to random walks}

The critical properties of the one-dimensional RTIM can be conveniently studied through a
mapping to a RW problem\cite{bigpaper} and a similar procedure works also for the random XX- and
XY-models\cite{ijr2000}. The method, which has also been used for inhomogeneous\cite{kjti}
and correlated disorder\cite{riegerigloi99}, is based on an exact expression of the surface
magnetization of the transverse-field Ising model with $L$ sites\cite{peschel,bigpaper}:
\be
m_s(L)=\left[1+\sum_{l=1}^{L-1}\prod_{j=1}^l \left({h_j \over J_j}\right)^2 \right]^{-1/2}\;,
\label{peschel}
\ee
where the last spin of the chain at $l=L$ is fixed to the state $|\sigma_L^x \rangle=|\uparrow \rangle$.
Before analysing Eq.(\ref{peschel}) we cite another simple relation\cite{itksz,bigpaper}, in which
the lowest excitation energy, $\epsilon(L)$, in a finite system with open boundary conditions is asymptotically
related to the surface magnetization in Eq.(\ref{peschel}):
\be
\epsilon(L) \sim m_s(L) \overline{m}_s(L) h_L \prod_{i=1}^{L-1} {h_i \over J_i}\;,
\label{epsilon}
\ee
provided $\epsilon(L)$ vanishes faster than $L^{-1}$. Here $\overline{m}_s(L)$ denotes the finite-size
surface magnetization at the other end of the chain and follows from the substitution $h_j/J_j \leftrightarrow
h_{L-j}/J_{L-j}$ in Eq.(\ref{peschel}).

Now we start to analyse the expression in Eq.(\ref{peschel}) and look for the possible values of
$m_s(L)$ using the discrete version of the distribution, $\pi(\Theta)$, fix $h_0=1$ and taking
the limit $\Lambda \to \infty$. It is easy to see that for this extreme distribution the products
in Eq.(\ref{peschel}),
$\prod_{j=1}^l \left({h_j / J_j}\right)^2$, take three different values: zero, one or infinity,
and for a given sample $m_s$ is zero, whenever any of the products is infinite, otherwise $m_s(L)=O(1)$.
To calculate the {\it average} surface magnetization one should collect the samples with $m_s(L)=O(1)$.
Here we use the RW-picture of Ref.\cite{bigpaper} and assign to each disorder configuration a
random walk, which starts at $t=0$ at position $y=0$ and takes at time $t=i$ a step of length
$\Theta_i$ with probability $\pi(\Theta_i)$.  Then for a disorder
configuration with a finite surface magnetization the corresponding RW stays
until $t=L$ steps at one side of its starting position,$y(t)>0,~t=1,2,\dots,L$,
in other words the RW has surviving character. As a consequence $[m_s(L)]_{\rm av}$
is proportional to the fraction of surviving RW-s, given by the surviving
probability, $P_{surv}(t)$ at $t=L$.

For a symmetric distribution, i.e. with
$p=q=1/2$ the corresponding RW-s have no drift,
whereas for the asymmetric case, $p \ne q$, there is an average bias
given by $\delta_W=q-p$, so that for $\delta_W>0(<0)$ the walk is drifted
towards (off) the adsorbing wall at $y=0$. The bias of the RW
is proportional to the control-parameter of the RTIM, $\delta$, as defined
in Eq.(\ref{delta}), thus the correspondence between RTIM and RW can be generally formulated as:
\be
\left[m_s(\delta,L)\right]_{\rm av} \sim \left. P_{surv}(\delta_w,t)\right|_{t=L},
\quad\delta \sim \delta_w.
\label{corresp}
\ee
Consequently from the persistence properties of L\'evy flights, which are summarized in the
Appendix, one can deduce the singular behavior of the average surface magnetization of the RTIM.

We start with the finite-size behavior at the critical point, $\delta=0$, which is given with the
correspondences in Eqs.(\ref{corresp}) and (\ref{symexp}) as
\be
\left[m_s(0,L)\right]_{\rm av} \sim L^{-x_m^s},\quad x_m^s=1/2\;.
\label{xms}
\ee
Thus the anomalous dimension of the average surface magnetization, $x_m^s$, does not depend on
the L\'evy index, $\alpha$, its value is the same as for the normal distribution in Table 1.

In the paramagnetic phase, $\delta>0$, the corresponding Riemann walk has an average drift towards the
adsorbing site. Consequently its surviving probability in Eq.(\ref{psurv+}) and thus the related
average surface magnetization of the RTIM has an exponentially decreasing behavior as a function of
the scaling variable, $\delta L^{1-1/\alpha}$, which is analogous to that in Eq.(\ref{longi}).
Consequently the characteristic length-scale in the problem, the average correlation length, $\xi$,
and the quantum control parameter, $\delta$, close to the critical point are related as in
Eq.(\ref{nu}), however with an $\alpha$-dependent exponent:
\be
\nu(\alpha)={\alpha \over \alpha -1}\;.
\label{nualpha}
\ee
Note that $\nu(\alpha)$ is divergent as $\alpha \to 1^+$, which is a consequence of the fact that the
first moment of the L\'evy distribution is also divergent in that limit. In the other limiting case,
$\alpha \to 2^-$, we recover the result for the normal distribution in Table 1.

In the ferromagnetic phase, $\delta<0$, the corresponding Riemann walk is drifted off the adsorbing
site and, as shown in the Appendix, the surviving probability approaches a finite value in the
large time limit. Consequently the average surface magnetization of the RTIM is also finite in the
ferromagnetic phase and for a small $|\delta|$ it behaves according to Eq.(\ref{psurv-}) as:
\be
\lim_{L \to \infty} \left[m_s(\delta,L)\right]_{\rm av} \sim |\delta|^{\beta_s},\quad \beta_s={\alpha
\over 2(\alpha -1)}\;.
\ee
Thus the scaling relation, $\beta_s=x_m^s \nu$, is satisfied.

Next, we turn to study the scaling behavior of the lowest excitation energy starting with the expression
in Eq.(\ref{epsilon}). Here we note that in a given sample the existence of a very small gap is
accompanied by the presence of local order. Thus in a sample with a low-energy excitation one can
find a strongly coupled domain (SCD) of size $l$, where the coupling distribution follows a
surviving walk character. Consequently in Eq.(\ref{epsilon}) $m_s=O(1)$ and $\overline{m}_s=O(1)$ and one
gets the estimate:
\be
\epsilon \sim \prod_{i=1}^{l-1} {h_i \over J_i} \sim \exp\left( - l_{tr} \overline{\ln(J/h)} \right)\;.
\label{epstr}
\ee
Here $l_{tr}$ measures the size of transverse fluctuations of the corresponding surviving L\'evy flight
of length $t=l$ and $\overline{\ln(J/h)}$ denotes an average value. At the critical point,
$\delta \sim \delta_w=0$, the surviving region of the walk and thus the SCD in the RTIM extends over
the volume of the sample, $l \sim L$, and from Eq.(\ref{transv}), $l_{tr} \sim L^{1/\alpha}$, so
that Eq.(\ref{epstr}) leads to:
\be
\ln \epsilon(L) \sim L^{\psi},\quad \psi=1/\alpha\,\quad \delta=0;.
\label{eps0}
\ee
The critical exponent, $\psi$ defined in Eq.(\ref{psi}) is a continuous function of the L\'evy index
which takes its value for the normal distribution in Table 1 when $\alpha \to 2^-$ .

In the paramagnetic phase, $\delta > 0$, there are still realizations with a small energy gap,
the scaling form of which in a finite system of size, $L$, can be estimated by the following
reasoning. For a L\'evy flight the probability of a large transverse fluctuation, $l_{tr}$,
is given from Eq.(\ref{largeu}) as $p(l_{tr},L) \sim L \left( l_{tr} \right)^{-(1+\alpha)}$, thus its
characteristic size can be estimated as
$l_{tr} \sim L^{1/(1+\alpha)}$ from the condition $p(l_{tr},L)=O(1)$.
Consequently the lowest gap has the scaling form:
\be
\ln \epsilon(L) \sim L^{1/(1+\alpha)},\quad \delta>0\;,
\label{eps+}
\ee
which implies a logarithmically broad gap distribution even in the Griffiths phase. This is in contrast
to the behavior with the normal distribution, where $l_{tr} \sim \ln L$\cite{bigpaper}, and the scaling
form of the gap is in a power-law form, $\epsilon(L) \sim L^{-z}$, where the dynamical exponent, $z$,
is a continuous function of the quantum control parameter.

In the remaining part of this subsection we discuss the probability distribution of the surface magnetization.
Let us remind that, at the critical point, the average surface magnetization is determined by the
so-called {\it rare events}, which are samples having $m_s=O(1)$. The typical samples, however, which are
represented by non-surviving random walks, have a vanishing surface order in the thermodynamic limit.
For a large but finite system of size $L$, $m_s(L)$ is dominated by the largest product, $\prod_j (h_j/J_j)$,
in Eq.(\ref{peschel}) so that
\be
\ln m_s(L) \sim \overline{\epsilon}(L) \sim -l_{tr} \overline{\ln(h/J)}
\label{lnms}
\ee
where $\overline{\epsilon}(L)$ is the value of the gap in the dual system, i.e. where the fields and the
couplings are interchanged, $h_i \leftrightarrow J_i$. Since at the critical point the system is self-dual
we have from Eq.(\ref{eps0}) $\ln m_s(L) \sim L^{\psi}$ and the appropriate scaling variable is
\be
P_L(\ln m_s)={1 \over L^{\psi}} \tilde{p}\left(\ln m_s \over L^{\psi} \right)\;.
\ee

\begin{figure}[ht]
\epsfxsize=8truecm
\begin{center}
\mbox{\epsfbox{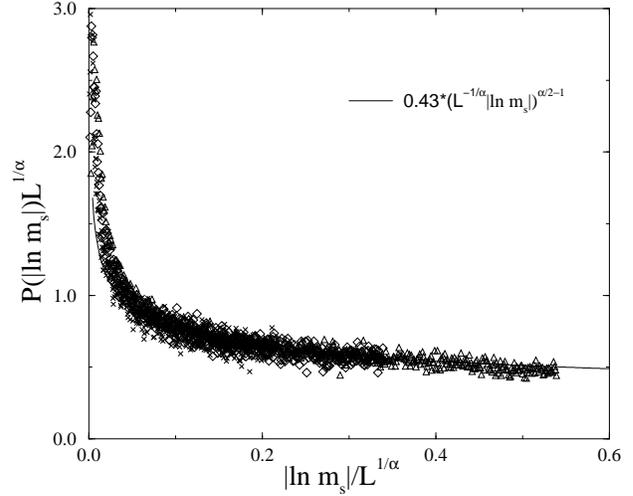}}
\end{center}
\caption{Rescaled distribution of the logarithmic surface magnetization for $L=512(\bigtriangleup),1024(\diamond)$ and
$2048(\circ)$ at
$\alpha=1.5$. The full line represents the scaling expression for small arguments, where the only
fitting parameter is the amplitude.}
\label{fig1} 
\end{figure}

Assuming that the scaling function behaves as $\tilde{p}(y) \sim |y|^a$ for small $|y|$,
then the average magnetization is given by $\left[ m_s \right]_{\rm av}=1/L^{\psi} \int {\rm d}m_s
\tilde{p}(\ln m_s / L^{\psi}) \sim L^{-(1+a) \psi}$, which implies $a=\alpha/2-1$, to recover the exponents
in Eqs.(\ref{xms}) and (\ref{eps0}). These relations have been checked through a numerical evaluation of
Eq.(\ref{peschel}) for 50000 samples of large systems with size up to $L=2048$. As seen in Fig.1 there is a very good
agreement between the scaling and numerical results.
In the paramagnetic phase, $\delta>0$, close to the critical point according to Eq.(\ref{lnms}) the typical surface magnetization
behaves asymptotically as
\be
\ln m_s(L) \sim -\delta L \sim -L/\xi_{typ}\;,
\ee
where the typical correlation length, $\xi_{typ}$, diverges at the critical point as:
\be
\xi_{typ} \sim |\delta|^{-\nu_{typ}},\quad \nu_{typ}=1\;.
\ee
Note that the characteristic exponent, $\nu_{typ}$, is independent of $\alpha$, but still satisfies
the scaling relation\cite{fisherA}, $\nu_{typ}=\nu(1-\psi)$.

\subsection{Bulk magnetization: a numerical renormalization group study}

The critical behavior of the bulk magnetization of the 1d RTIM is not related to the properties
of a homogeneous RW, but it can be calculated from the so called average persistence properties
of a Sinai walk, i.e. a random walk in a random environment\cite{riegerigloi99b}.
This procedure has already been
used to calculate the magnetization scaling dimension, $x_m$, of the RTIM with correlated
disorder\cite{riegerigloi99}.
Here we use another numerical method to study $x_m$ for L\'evy-type disorder, which method is based on
a numerical implementation of the Ma-Dasgupta-Hu RG procedure. The RG-equations\cite{fisher} and
their numerical use are well documented in the literature\cite{2drg,lkir,ki}. Here we use the
finite-size version of the method\cite{lkir}, in which we start with a large finite ring of size $L$ 
with random couplings and perform the decimation procedure until the last spin. The last
log-coupling (transverse field) sets the log-energy scale, $\Gamma=-\log(\Omega/\Omega_0)$, which
at the critical point scales from Eq.(\ref{psi})
as $\Gamma \sim L^{\psi}$, whereas the critical cluster
moment in Eq.(\ref{phi}), associated with the last remaining cluster,
scales as $\mu \sim L^{\phi \psi}$. Repeating the calculation for several realizations of the
disorder the critical exponents can be deduced from the appropriate scaling functions.

We start with an analysis of the log-energy distribution, which should be a function of the
scaling variable $\Gamma/L^{\psi}$ for different finite systems. As shown in Fig.2 there is
an excellent data collapse
using our analytical result $\psi=1/\alpha$ as given in Eq.(\ref{eps0}).
\begin{figure}[ht]
\epsfxsize=8truecm
\begin{center}
\mbox{\epsfbox{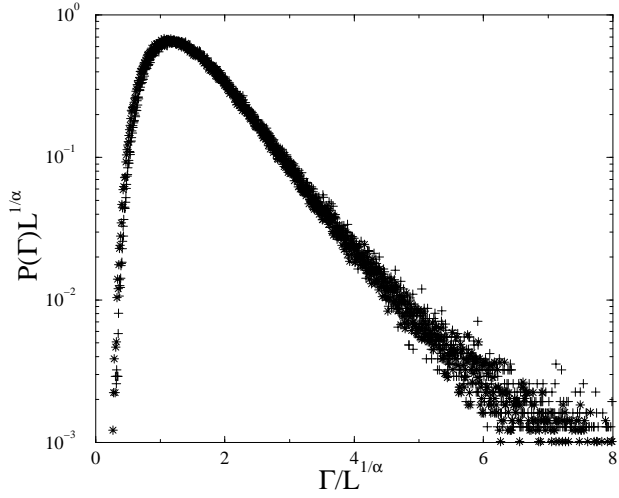}}
\end{center}
\caption{\label{fig2} Rescaled distribution of the logarithmic energy scale, $\Gamma$, for different
finite lengths, $L=32(\diamond),64(+)$ and $128(*)$ at $\alpha=1.5$. The best collapse of the data is obtained by
the analytical result $\psi=1/\alpha$.}
\end{figure}
Next, we analyze the distribution of the cluster moment $\mu$, which is a function of the
scaling variable $\mu/L^{\phi \psi}$, from which the magnetization scaling dimension, $x_m$,
follows through Eq.(\ref{xm}). From the optimal data collapse, as presented in Fig.3, we
obtain $x_m=0.22$ for $\alpha=1.5$, which is definitely larger than that of the
system with normal disorder, as given in Table 1. Repeating the calculation for the average magnetization,
$m(\delta)=[\mu/L]_{\rm av}$, the appropriate scaling variables outside the critical point are
$mL^{x_m}$ and $L \delta^{\nu}$. As shown in Fig.4 in terms of the scaled variables we
obtain a very good data collapse, using $x_m$ from Fig.3 and the
analytical expression in Eq.(\ref{nualpha}) for $\nu$. 

\begin{figure}[ht]
\epsfxsize=8truecm
\begin{center}
\mbox{\epsfbox{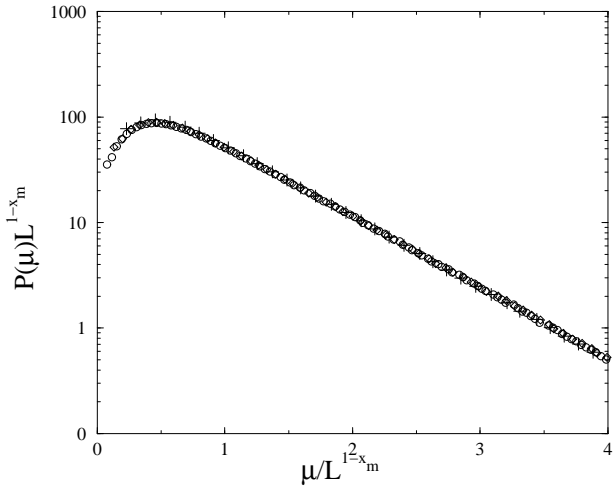}}
\end{center}
\caption{\label{fig3} Rescaled distribution of the cluster moment for different
finite lengths, $L=32(+),64(\diamond)$ and $128(\circ)$ at $\alpha=1.5$. 
The optimal collapse of the data gives $x_m=0.22$.}
\end{figure}

\begin{figure}[ht]
\epsfxsize=8truecm
\begin{center}
\mbox{\epsfbox{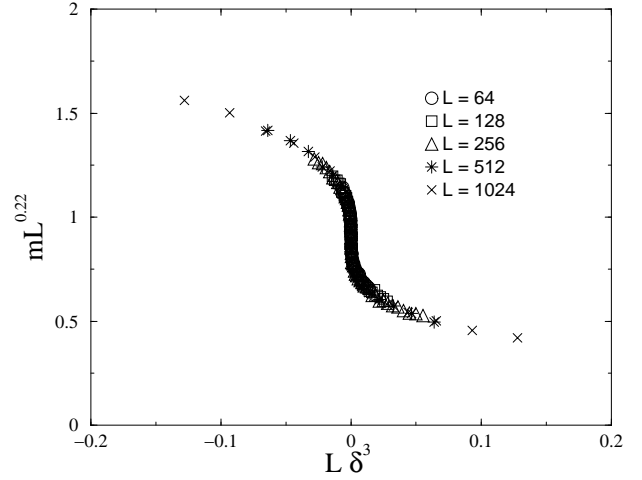}}
\end{center}
\caption{\label{fig4} Scaling plot of the average bulk magnetization for different finite
systems at $\alpha=1.5$.}
\end{figure}

For a systematic study of the bulk magnetization scaling index we have determined the
average critical magnetization for different finite systems and then calculated an effective
exponent, $x_m(L)$, by two-point fit comparing $m(L)$ and $m(2L)$. As shown in Fig.5
the effective exponents have a weak size dependence, so that one can have
an accurate extrapolated value for $L \to \infty$, which clearly depends on the value of
the  L\'evy index, $1<\alpha<2$. These extrapolated magnetization exponents are presented in
Fig. 6, where for $\alpha>2~~x_m$ is expected to be $\alpha$ independent, however the corrections
to scaling are strong, especially around the cross-over value $\alpha_c=2$. At this point we note
that in parallel to the RTIM we have also calculated the magnetization scaling dimension of the
random quantum Potts chain for $q=3$.

\begin{figure}[ht]
\epsfxsize=8truecm
\begin{center}
\mbox{\epsfbox{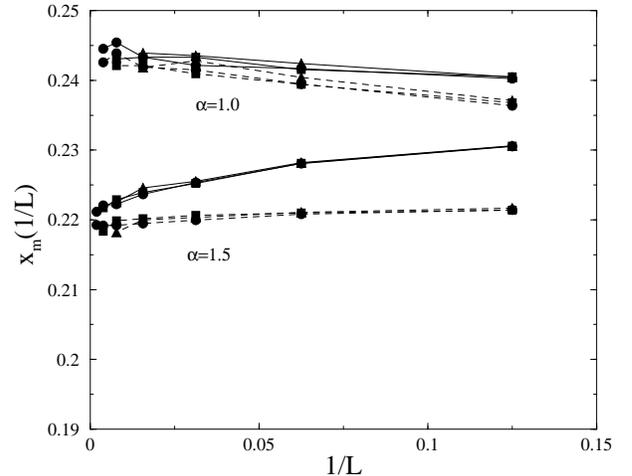}}
\end{center}
\caption{\label{fig5} Effective magnetic exponents for $\alpha=1.5$ and $\alpha=1.$ The
full lines connecting the data for the RTIM are guide to the eye. For a comparison the
same quantities for the $q=3$-state random quantum Potts chain are also presented, where
the data points are connected by dashed lines.}
\end{figure}

As can be seen in Fig. 5 the extrapolated exponents of
the Ising and Potts chains are very probably the same for the same value of the L\'evy index,
$\alpha$. This result completes the universality of the two models
as obtained before analytically for normal disorder\cite{senthil}.

We close this section by a study of the average magnetization in the ferromagnetic phase, $\delta<0$,
where the average cluster moment scales with the size of the system, $L$.
As a consequence the average magnetization approaches a finite limit as $L \to \infty$, which
close to the critical point behaves asymptotically as $m(\delta) \sim |\delta|^{\beta}$. As shown in Fig.7
the results for different finite systems converge to a power-law  form where the critical exponent,
$\beta$, satisfies the scaling relation, $\beta=x_m \nu$.

\begin{figure}[ht]
\epsfxsize=8truecm
\begin{center}
\mbox{\epsfbox{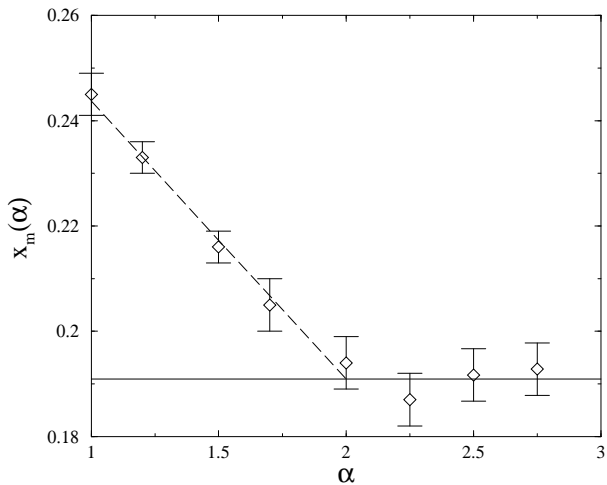}}
\end{center}
\caption{\label{fig6} Bulk magnetization exponent for different values of $\alpha$. The full line
represents $x_m$ for normal disorder, the dashed line is guide to the eye.}
\end{figure}

\begin{figure}[ht]
\epsfxsize=8truecm
\begin{center}
\mbox{\epsfbox{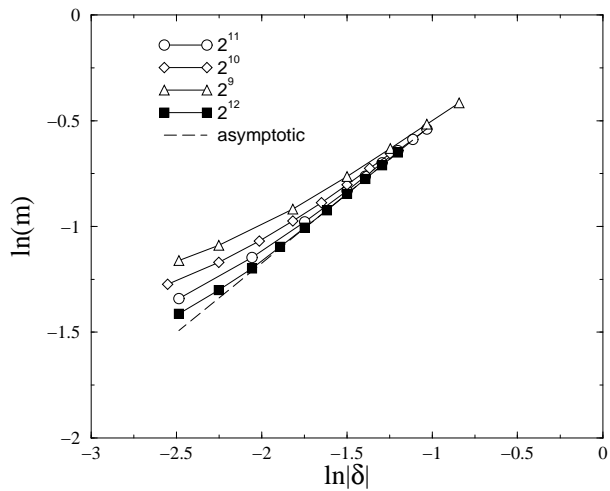}}
\end{center}
\caption{\label{fig7} Average bulk magnetization in the ordered phase 
as a function of the quantum control parameter in a log-log plot ($\alpha=1.5$). For small $\delta$
the finite size results converge to the dashed line with a slope of $\beta=0.66$.}
\end{figure}

\section{The two-dimensional problem}

In 2d the critical properties of the RTIM with normal disorder is controlled
by an IRFP, so that the Ma-Dasgupta-Hu
RG procedure provides asymptotically exact critical properties. We expect that these properies of the
RTIM stay valid for broad distributions, too, and we study the 2d problem using the RG method.
There are, however, several limitations for the numerical
implementation of the method. First, as in 1d, the systems under consideration have a finite
spatial extent and the number of disorder realizations which can be studied is also finite. In 2d an
extra problem is caused by the fact that the topology of the lattice is not invariant under the
RG transformation and as a result new couplings between remote
spins are generated.

To avoid these type of difficulties
a truncation procedure has been introduced in Ref\cite{2drg}, such that a class of generated
interactions, which are expected to cause very small errors, are neglected. In
Ref.[\cite{lkir}], where the finite-size version of the RG method is used, all the generated
interactions were kept for moderately large finite systems. In Ref.[\cite{ki}] a selection
condition in the finite-size RG method has been introduced by showing that many of the
generated new couplings are ``dead'' in the sense that they are not
decimated out in later steps, so that they can be omitted without causing any error in the
renormalization. With this observation one could consider larger finite systems and at the
same time the necessary computational time has considerably reduced.

In the present paper we apply the finite-size RG method supplemented by the selection condition.
In this way we could treat systems on the square lattice with linear size up to $L=128$ and we
considered typically 10000 realizations and some 1000 samples for the largest systems. 

In 2d the position of the critical point of the system is not known by
self-duality, therefore we used the following numerical procedure for its determination.
First we calculated the average magnetization,
$m(L,h_0)$, in a finite systems of size $L$ and then defined the scaling function,
$g_L(h_0)$, as the ratio:
\be
{m(L,h_0) \over m(L/2,h_0)}=g_L(h_0)\;.
\ee
In the ferromagnetic phase, $h_0<h_c$, in the thermodynamic limit $m(L,h_0)$ does not depend on $L$,
consequently $\lim_{L \to \infty} g_L(h_0)=1$. On the other hand in the paramagnetic phase, $h_0>h_c$,
using the example of the surface magnetization of the 1d RTIM in Eq.(\ref{peschel}), $m(L,h_0)$ is  
exponentially small in $L$, thus the scaling function has a vanishing limiting value.
In between at the critical point, $h_0=h_c$, where $m(L,h_c) \sim
L^{-x_m}$ we have a finite limiting value $\lim_{L \to \infty} g_L(h_c)=2^{-x_m}$. Consequently
calculating $g_L(h_0)$ for a series of sizes the position of the limiting crossing points defines
the critical point of the system, whereas the abscissas of the crossing points are related to the
magnetization scaling dimension, $x_m$. This procedure is illustrated in Fig. 8 on the example of
the 1d model with $\alpha=1.5$, where the previously calculated critical properties of the
system are accurately reobtained.

\begin{figure}[ht]
\epsfxsize=8truecm
\begin{center}
\mbox{\epsfbox{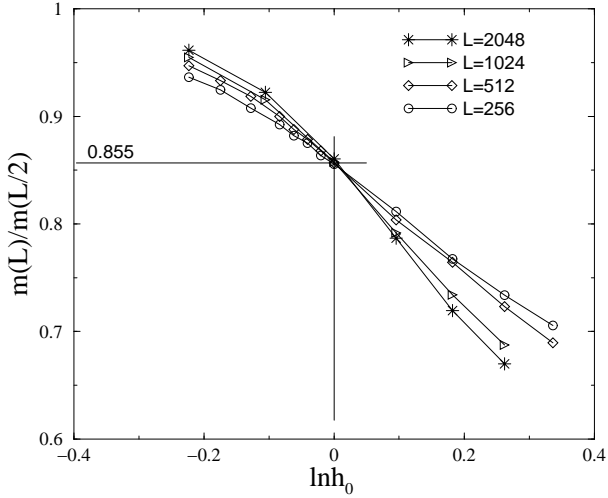}}
\end{center}
\caption{\label{fig8} Finite-size estimates of the critical point and the magnetization scaling
dimension of the 1d model with $\alpha=1.5$.}
\end{figure}

\begin{figure}[ht]
\epsfxsize=8truecm
\begin{center}
\mbox{\epsfbox{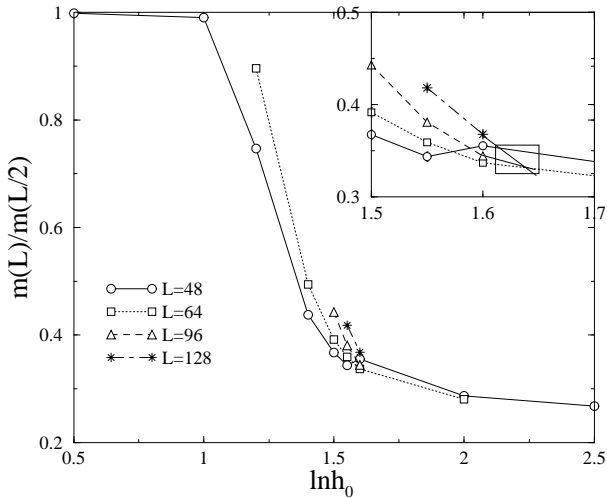}}
\end{center}
\caption{\label{fig9} The same as in Fig. 8 for the 2d model with $\alpha=3$.}
\end{figure}

In 2d a similar plot of the scaling functions for $\alpha=3$ is shown in Fig. 9, where the crossing
point is obtained with a reasonably small error. Repeating this calculation for various
$\alpha$-s we have obtained estimates for the critical points and the magnetization exponents, which
are presented in Fig. 10. As seen in this figure $x_m$ is monotonously decreasing with $\alpha<\alpha_c$,
whereas for $\alpha>\alpha_c~~x_m$ stays approximately constant and this value corresponds, within
the error of estimates, to that of the 2d RTIM with normal disorder, as given in Table 1.
>From Fig. 10 the cross-over value can be estimated as $\alpha_c \approx 4.5$.
The magnetization scaling dimension, $x_m$, or more precisely $\phi \psi=2-x_m$, is also related
to the finite-size behavior of the average cluster moment at the critical point: $[\mu_L]_{\rm av}
\sim L^{\phi \psi}$, which is illustrated in Fig.11 for $\alpha=3$. From the slope of the curve
in a log-log plot one obtains an estimate of $\phi \psi=.48$, which is in agreement with the
results shown in Fig. 10. Similar agreement is found for other values of $\alpha$, too.

\begin{figure}[ht]
\epsfxsize=8truecm
\begin{center}
\mbox{\epsfbox{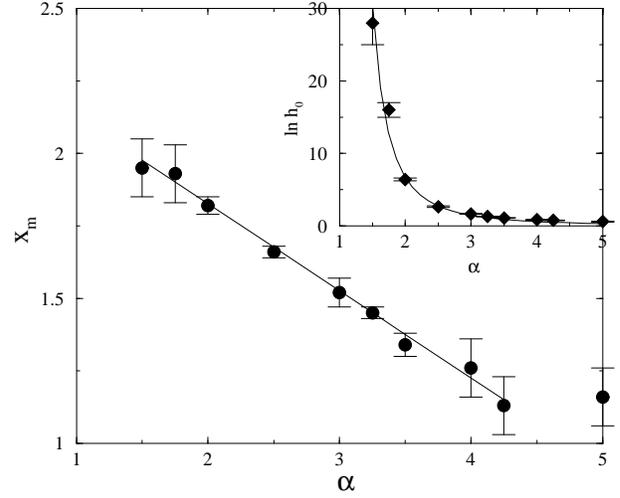}}
\end{center}
\caption{\label{fig10} Magnetization scaling dimensions in 2d for different values of the Levy-index.
In the inset the estimated critical points are presented. The line connecting the data points is a
guide to the eye. }
\end{figure}

\begin{figure}[ht]
\epsfxsize=8truecm
\begin{center}
\mbox{\epsfbox{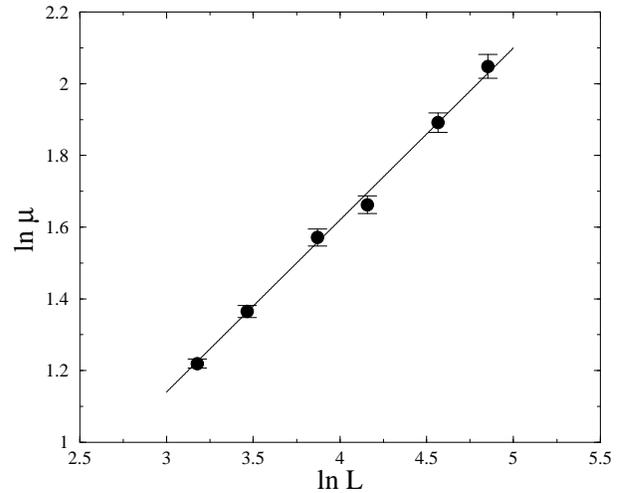}}
\end{center}
\caption{\label{fig11} Average critical cluster moment as a function of size in a log-log plot
($\alpha=3$). The slope of the line corresponds to $\phi \psi=0.48$.}
\end{figure}

Our final investigation concerns the scaling behavior of the distribution function of the
log-energy-scale, $P_L(\Gamma)$, at the critical point. As in the 1d problem the critical exponent,
$\psi$, can be obtained from an optimal data collapse in terms of the scaling variable
$\Gamma/L^{\psi}$. In Fig. 12 we present for $\alpha=3$ the rescaled accumulated probability
distribution function, where a satisfactory data collapse is obtained for $\psi=0.8$. Similar
estimates for other $\alpha$-s are collected in Fig. 13. As for the $x_m$ exponent $\psi$ is a
monotonously decreasing function for $\alpha<\alpha_c \approx 4.5$, whereas for $\alpha>\alpha_c$ it is
approximately constant and this value corresponds to that of the model with normal disorder
as given in Table 1.

\begin{figure}[ht]
\epsfxsize=8truecm
\begin{center}
\mbox{\epsfbox{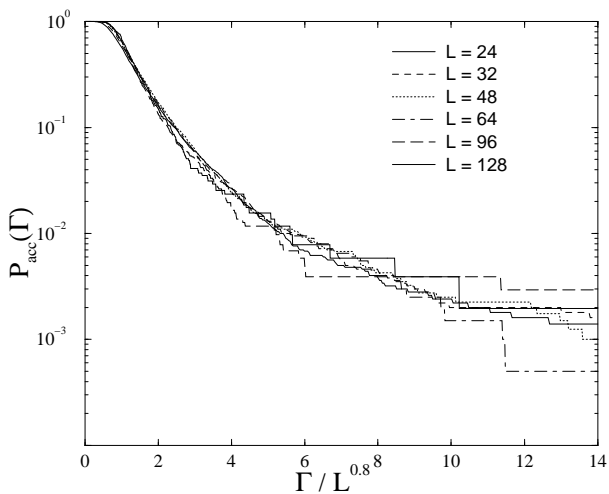}}
\end{center}
\caption{\label{fig12} Rescaled accumulated probability distribution function of the log-energy-scale
at $\alpha=3$. Satisfactory data collapse is obtained with the exponent $\psi=0.80$.}
\end{figure}

\begin{figure}[ht]
\epsfxsize=8truecm
\begin{center}
\mbox{\epsfbox{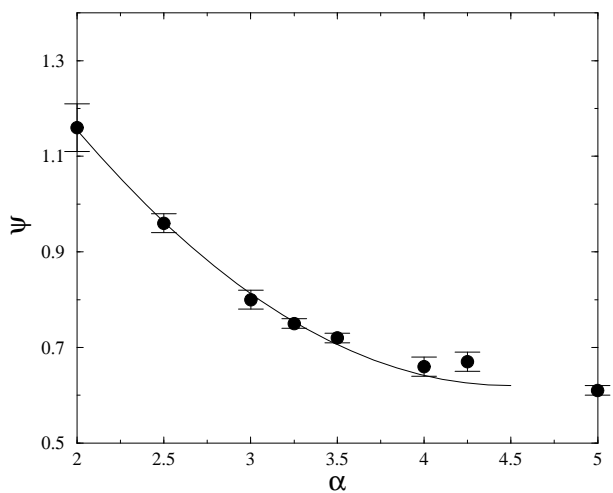}}
\end{center}
\caption{\label{fig13} The critical exponent $\psi$ as a function of the L\'evy index for the
2d model. The line is a guide to the eye.}
\end{figure}

\section{Summary}

In this paper the effect of a broad disorder distribution on the critical behavior of the RTIM
is studied in one and two space dimensions. The broadness of the
disorder distribution becomes relevant, if the L\'evy index is lowered below a critical
value, $\alpha_c$. In the region of $1<\alpha<\alpha_c$ the critical exponents of the IRFP-s
are continuous functions of $\alpha$ and for $\alpha>\alpha_c$ they are the same as in
the model with normal disorder. 

In 1d we obtained $\alpha_c=2$, in close analogy with random walks, where the central limit
theorem is valid for $\alpha>2$. This analogy is more than a simple coincidence, since the 1d RW and the
quantum Ising spin chain are related through an exact mapping\cite{itr99}, which then has the same
requirement for the relevance-irrelevance conditions. In 2d, where such type of
mapping does not exist, the limiting value is found to be approximately around 
$\alpha_c \approx 4.5$, thus in the region of
$2<\alpha<\alpha_c$ the broadness of disorder is relevant for the RTIM, whereas it is irrelevant
for the RW.

Outside the critical point, in the so-called Griffiths-phase\cite{griffiths} some physical
quantities of random quantum Ising magnets (linear and non-linear susceptibility,
autocorrelations, etc.) are still singular and in the presence of normal disorder
these Griffiths-McCoy singularities are characterised by the dynamical exponent, $z(\delta)$,
which is a function of the distance of the critical point. For broad distribution of the
disorder the probability of rare events is enhanced, consequently the Griffiths-McCoy singularities
become stronger. In 1d we have shown by scaling arguments leading to Eq.(\ref{eps+}) that the
dynamical exponent is formally infinite in the whole Griffiths-phase.

Considering other magnetic models we expect that in 1d the critical properties of the random XX,
XXX and q-state quantum Potts models are controlled by the same IRFP as that of the RTIM and this
is parametrized by the L\'evy index, $\alpha$. 
This result for the XX-model is a consequence of
a known mapping,\cite{fisherxx,ijr2000} whereas for the two other models the RG decimation equations
are asymptotically equivalent to that of the RTIM\cite{fisherxx,senthil}. This equivalence has been
explicitly tested here for the Potts model with $q=3$.
On the other hand in two dimensions such an equivalence can be
expected only for the random quantum Potts model, since the random XX- and XXX-models have no
IRFP-s in higher dimensions\cite{2drg,ijr2000}.

\begin{acknowledgement}
We are grateful to L. Turban for useful discussions and for a critical reading
of the manuscript. This work has been
supported by a German-Hungarian exchange program (DAAD-M\"OB), by the Hungarian
National Research Fund under grant No OTKA
TO23642, TO25139,  MO28418  and by the Ministery of Education under grant No. FKFP 0596/1999.
Numerical calculations are partially performed on the Cray-T3E at Forschungszentrum J\"ulich.
The Laboratoire de Physique des Materiaux is Unit\'e Mixte de Recherche CNRS No 7556.
\end{acknowledgement}

\Appendix{Surviving probability of L\'evy flights}

Consider the following sum
\be
S_n=\sum_{j=1}^n x_j\;,
\ee
where the independent random variables, $x_j$, follow the same broad probability distribution,
$\pi(x)$, which asymptotically behaves as
\be
\pi(x) \simeq_{x\to \infty} p x^{-(1+\alpha)}\quad ,\qquad\pi(x) \simeq_{x\to -\infty} q |x|^{-(1+\alpha)}\quad \;,
\ee
with $1<\alpha<2$
and we are interested in the probability distribution of $S_n$, $p(S,n)$, for large $n$.
According to exact results\cite{kolmogorov} there exists a limit distribution, $\tilde{p}(u)\rm{d}u$, in term
of the variable, $u=S_n/l_n-c_n$, as $n \to \infty$. Here the normalization:
\be
l_n=n^{1/\alpha}\;,
\label{transv}
\ee
is the transverse fluctuation of the walk, if we interpret $n=t$ as the (discrete) time and $S_{n=t}$ as
the position of the walker in the transverse direction. The second normalization is given by
\be
c_n=-n^{1-1/\alpha} \delta_w\;,
\label{longi}
\ee
where with $\delta_w=-\langle x \rangle$ we define the bias of the walk. For a small $\delta_w$
one gets from the combination in Eq.(\ref{longi}) the scaling relation between time and bias as:
\be
t \sim |\delta_w|^{-\nu(\alpha)},\quad \nu(\alpha)={\alpha \over \alpha-1}\;.
\label{nuwalk}
\ee
For a symmetric distribution, when $\pi(x)$ is an even function, thus $p=q$ and $\delta_w=0$, we
have for the limit distribution:
\be
\tilde{p}(u)=L_{\alpha,0}(u)={1 \over 2 \pi} \int_{-\infty}^{\infty} e^{iku-|k|^{\alpha}} {\rm d}k\;,
\ee
which has an expansion around $u=0$
\be
L_{\alpha,0}(u)={1 \over \pi \alpha} \sum_{k=0}^{\infty} (-1)^k {u^{2k} \over (2k)!} \Gamma\left({2k+1 \over \alpha}
\right)\;
\label{smallu}
\ee
and for large $u$ it is asymptotically given by
\be
L_{\alpha,0}(u)={1 \over \pi} u^{-(1+\alpha)} \Gamma(1+\alpha) \sin(\pi \alpha/2)\;,
\label{largeu}
\ee
where $\Gamma(x)$ denotes the gamma function.

Consider next the surviving probability, $P_{surv}(t,\delta_w)$, which is given by the fraction of
those walks, which have not crossed the starting position until $t=n$, thus $S_i>0$ for $i=1,2,\dots,n$.
For a biased walk, with $0<|\delta_w|\ll 1$, the asymptotic behavior of $P_{surv}(n,\delta_w)$ is equivalent
to that of a symmetric walk ($\delta_w=0$) but with a moving adsorbing boundary site, which has a constant
velocity of $v=\delta_w$. For this event, with $S_i>vi$ for $i=1,2,\dots,n$, the surviving probability
is denoted by $F(n,v)$, whereas the probability for $S_n>vn$, irrespectively from the previous steps, is
denoted by $P(n,v)$ and the latter is given by:
\be
P(n,v)=\int_{nv}^{\infty} p(S,n) {\rm d} S\;.
\label{pnv}
\ee
Between the generating functions:
\beqn
F(z,v)&=&\sum_{n \ge 0} F(n,v) z^n\nonumber\\
P(z,v)&=&\sum_{n \ge 1} {P(n,v) \over n}z^n\;
\eeqn
there is a useful relation due to Sparre Andersen\cite{andersen}:
\be
F(z,v)=\exp\left[P(z,v)\right]\;,
\ee
which has been used recently in Ref\cite{luck}.

In the zero velocity case, $v=0$, which is equivalent to the symmetric walk with $\delta_w=0$, we
have $P(n,0)$=1/2. Consequently $P(z,0)=-{1 \over 2} \ln(1-z)$ and $F(z,0)=(1-z)^{-1/2}$, from which
one obtains for the final asymptotic result:
\be
P_{surv}(t,0)=\left. F(n,0)\right|_{n=t} \sim t^{-\theta},\quad\theta=1/2\;.
\label{symexp}
\ee
Note that the persistence exponent, $\theta=1/2$, is independent of the form of a symmetric
probability distribution, $\pi(x)$, thus it does not depend on the L\'evy index, $\alpha$.

For $v>0$, i.e. when the allowed region of the particle shrinks in time the correction to
$P(n,0)=1/2$ has the functional form, $P(n,v)=1/2-g(\tilde{c})$, with $\tilde{c}=v n^{1-1/{\alpha}}$.
Evaluating Eq.(\ref{pnv}) with Eq.(\ref{smallu}) one gets in leading order,
$P(n,v)={1 \over 2} - \tilde{c} A(\alpha) +O(\tilde{c}^3)$,
with $A(\alpha)=\Gamma(1+1/\alpha)/\pi$. Then $P(z,v)-P(z,0)\simeq A(\alpha) v \sum_{n \ge 1} z^n n^{-1/\alpha}$
is singular around $z \to 1^-$ as $\sim (1-z)^{-(1-1/\alpha)}$, consequently
\be
F(z,v)\simeq (1-z)^{-1/2} \exp\left[-A(\alpha) v (1-z)^{-(1-1/\alpha)} \right]\;,
\ee
in leading order and close to $z=1^{-}$. Here the second factor gives the more singular contribution
to the surviving probability, which is in an exponential form:
\begin{eqnarray}
P_{surv}(t,\delta_w) &\sim &\left.F(n,v)\right|_{v=\delta_w,n=t} \nonumber\\
&\sim& t^{-1/2} \exp\left[ -{\rm const} \delta_w
t^{1-1/\alpha} \right]\;.
\label{psurv+}
\end{eqnarray}

For $v<0$, i.e. when the allowed region of the particle increases in time we consider the large $|v|$ limit
and write Eq.(\ref{pnv}) with Eq.(\ref{largeu}) as $P(n,v) \simeq 1 - B(\alpha) \tilde{c}^{-\alpha}+O(\tilde{c}^
{-3\alpha})$ with $B(\alpha)=\Gamma(1+\alpha) \sin(\pi \alpha/2)/\pi \alpha$. Then, in the large $|v|$ limit
$P(z,v)=-\ln(1-z)-B(\alpha) |v|^{-\alpha} \sum_{n\ge 1} z^n n^{-\alpha}$, where the second term is convergent
even at $z=1$. As a consequence the surviving probability remains finite as $n \to \infty$ and we have the
result, $F(n,v)\simeq 1 - {\rm const} |v|^{-\alpha}$, for $|v| \gg 1$. For a small velocity, $0<|v|\ll 1$,
we can estimate $F(n,v)$ by the following reasoning. After $n=n_c$ steps the distance of the adsorbing
site from the starting point, $y_s=v n_c$, will exceed the size of transverse fluctuations of the walk in
Eq(\ref{transv}), $l_{tr} \sim n_c^{1/\alpha}$, with $n_c \sim |v|^{-\nu(\alpha)}$. Then the walker which
has survived until $n_c$-steps with a probability of $n_c^{-1/2}$, will survive in the following
steps with probability $O(1)$. Consequently
\be
\lim_{t \to \infty} P_{surv}(n,\delta_w) \sim \lim_{n \to \infty} \left. F(n,v) \right|_{v=\delta_w, n=t}
\sim |\delta_w|^{\nu(\alpha)/2}\;.
\label{psurv-}
\ee

\end{document}